\documentclass[reprint,superscriptaddress,amsmath,amssymb,aps,pra,usenames,dvipsnames]{revtex4-1}

\usepackage{graphicx}% Include figure files
\usepackage{dcolumn}% Align table columns on decimal point
\usepackage{bm}% bold math
\usepackage{xcolor}
\usepackage{MnSymbol,wasysym}
\usepackage{amsmath}
\usepackage[normalem]{ulem} %it allows \sout

\begin{document}

\preprint{AIP/123-QED}

\title{ Intrinsic mitigation of the after-gate attack in quantum key distribution through fast-gated delayed detection }
%\title{Exploiting detector imperfection to mitigate the after-gate attack in quantum key distribution}
%Accurate modelling of a self-differencing avalanche photodiode under bright optical illumination]{Accurate modelling of a self-differencing avalanche photodiode under bright optical illumination}

\author{A. Koehler-Sidki}
 \affiliation{
Toshiba Research Europe Ltd, Cambridge Research Laboratory, 208 Cambridge Science Park, Milton Road, Cambridge, CB4 0GZ, United Kingdom%\\This line break forced with \textbackslash\textbackslash
}%
 \affiliation{%
Engineering Department, University of Cambridge, 9 J. J. Thomson Avenue, Cambridge CB3 0FA,
United Kingdom%\\This line break forced% with \\
}%
%\email{Alex.Koehler-Sidki@crl.toshiba.co.uk}

\author{J. F. Dynes}%
\email {james.dynes@crl.toshiba.co.uk}
\author{A. Martinez}%
\author{M. Lucamarini}%

\affiliation{
Toshiba Research Europe Ltd, Cambridge Research Laboratory, 208 Cambridge Science Park, Milton Road, Cambridge, CB4 0GZ, United Kingdom%\\This line break forced with \textbackslash\textbackslash
}%
\author{G. L. Roberts}
\affiliation{
Toshiba Research Europe Ltd, Cambridge Research Laboratory, 208 Cambridge Science Park, Milton Road, Cambridge, CB4 0GZ, United Kingdom%\\This line break forced with \textbackslash\textbackslash
}%
 \affiliation{%
Engineering Department, University of Cambridge, 9 J. J. Thomson Avenue, Cambridge CB3 0FA,
United Kingdom%\\This line break forced% with \\
}%
\author{A. W. Sharpe}
\affiliation{
Toshiba Research Europe Ltd, Cambridge Research Laboratory, 208 Cambridge Science Park, Milton Road, Cambridge, CB4 0GZ, United Kingdom}
%\author{S. J. Savory}
%\affiliation{%
%Engineering Department, University of Cambridge, 9 J. J. Thomson Avenue, Cambridge CB3 0FA,
%United Kingdom%\\This line break forced% with \\
%}%
\author{Z.~L.~Yuan}
\author{A. J. Shields}
\affiliation{
Toshiba Research Europe Ltd, Cambridge Research Laboratory, 208 Cambridge Science Park, Milton Road, Cambridge, CB4 0GZ, United Kingdom%\\This line break forced% with \\
}%

\date{\today}% It is always \today, today,

\begin{abstract}
 \noindent The information theoretic security promised by quantum key distribution (QKD) holds as long as the assumptions in the theoretical model match the parameters in the physical implementation.
The superlinear behaviour of sensitive single-photon detectors represents one such mismatch and can pave the way to powerful attacks hindering the security of QKD systems, a prominent example being the after-gate attack.
A longstanding tenet is that trapped carriers causing delayed detection can help mitigate this attack, but despite intensive scrutiny, it remains largely unproven.
Here we approach this problem from a physical perspective and find new evidence to support a detector's secure response.
We experimentally investigate two different carrier trapping mechanisms causing delayed detection in fast-gated semiconductor avalanche photodiodes, one arising from the multiplication layer, the other from the heterojunction interface between absorption and charge layers.
The release of trapped carriers increases the quantum bit error rate measured under the after-gate attack above the typical QKD security threshold, thus favouring the detector's inherent security. This represents a significant step to avert quantum hacking of QKD systems.
\end{abstract}

\pacs{Valid PACS appear here}% PACS, the Physics and Astronomy
%Use showkeys class option if keyword
\maketitle

\noindent Quantum key distribution (QKD) promises secure distribution of cryptographic digital keys \cite{BB14}, spurring significant development of the technology. This has rapidly matured and is now stepping out of the laboratory and into deployment in optical fibre networks \cite{peev_secoqc_2009,sasaki11,Dynes12,Mao:18,dixon_2017,Sun18,Bunandar18}. Contributing to its maturity, a great deal of research has been devoted to quantum hacking \cite{makarov_*_faked_2005,sauge_controlling_2011,vakhitov_tha_2011,gisin_tha_2006,jiang_intrinsic_2013}, which identifies imperfections of QKD components from their theoretical models and evaluate their implications for QKD security. Best-practice criteria and countermeasures can then be developed \cite{yuan2010avoiding,yuan_resilience_2011,Lydersen_bitmap_11,lucamarini_tha_2015,Lim15,daSilva15,Koehler18,KoehlerSidkiIM18} to reinforce the identified weak components and reclaim implementation security.

Due to their exposure to the quantum channel, single photon detectors in QKD systems have been subjected to most hacking attacks in the past decade \cite{lydersen_hacking_2010,Makarov16,Pinheiro18}. Weak detectors have been demonstrated to be under full control of an eavesdropper (Eve), resulting in a collapse of security \cite{gerhardt_full-field_2011}. Detector loopholes can be completely closed by novel protocols that achieve measurement-device independent security \cite{lo12,Braunstein12,LYDS18}. However, these protocols require an intermediate relay and therefore their deployment in the network is unfavorably complex when compared with standard point-to-point QKD links. A solution to regain detector security is thus highly desirable for relayless QKD links.
%Indeed, protocol-level attempts have been made, but without much success.

Single photon detectors based on semiconductor InGaAs avalanche photodiodes (APDs) serve the majority of links in existing QKD networks \cite{peev_secoqc_2009,sasaki11,Dynes12,Mao:18,dixon_2017,Sun18}, because they operate at temperatures that are easily within reach of thermo-electric cooling \cite{comandar_gigahertz-gated_2015_aip} or even room temperature \cite{comandar_rmtemp_2014}.
The state-of-the-art systems can offer a key rate exceeding 10 Mb/s \cite{Yuan18} and operate over 200 km fiber \cite{Boaron18}.

Attacks on InGaAs APDs have revealed their vulnerabilities, most of which have been dealt with because Eve's attack either changes the detector characteristics or produces a detectable fingerprint. However, as a special class of faked state attack \cite{makarov_*_faked_2005}, the faint after-gate attack \cite{Lydersen11} remains an open threat.  This is because detectors under such attack will maintain their single-photon sensitivity and will not produce a massive photocurrent \cite{Qian18} as in bright illumination attacks.

When a photon is absorbed by an InGaAs APD it generates an electron-hole pair.
The hole can then become trapped in defects or at barriers and is released with a certain probability related to the characteristic time constant of the trap. As opposed to trapped carriers arising from macroscopic avalanches, whose lifetimes are on the order of microseconds, the trapped hole lifetime at the material interface is on the sub-nanosecond order. Such trapping, therefore, does not have an effect in MHz-gated detectors \cite{Lydersen11,Qian18}. However, under GHz gating, the trapping time becomes comparable with the the detector gating period and the release of such carriers in subsequent gates can result in substantial amounts of delayed detection events. This could provide a means to mitigate the faint after-gate attack. Hence, it is natural to look at fast GHz-gated APDs \cite{comandar_gigahertz-gated_2015_aip} as a potential countermeasure to this attack.
So far, however, there is no study supporting this conjecture.
Earlier investigations have largely been concerned with MHz-gated detectors, where the time between gates is significantly larger than the decay time of trapped carriers.
Furthermore, the analysis of the  Quantum Bit Error Rate (QBER) has previously focused solely on Eve's target gate \cite{Lydersen11,Qian18},  due to the contribution from delayed detection events being negligible.
%As such, previous studies have not been able to exploit the delayed detection events studied in this paper.

In this work, we investigate two sources of carrier trapping in fast-gated InGaAs APDs, one from the multiplication layer and the other from the hetero-interface between the two materials, and find that both cause a non-negligible delayed detection probability.

This previously perceived drawback of single photon InGaAs APDs can be used to detect an after-gate attack.
The delayed photodetection introduces an increase in the QBER  of the QKD system that unveils the attack, thus promoting fast-gated devices as a means of mitigating this potential vulnerability.
In addition to that, we show that the amount of induced QBER in Eve's absence is not excessive and still allows for efficient QKD operation if the appropriate gating frequency is chosen.

\begin{figure}[htbp!]
  \centering
  \includegraphics[width=0.4\textwidth]{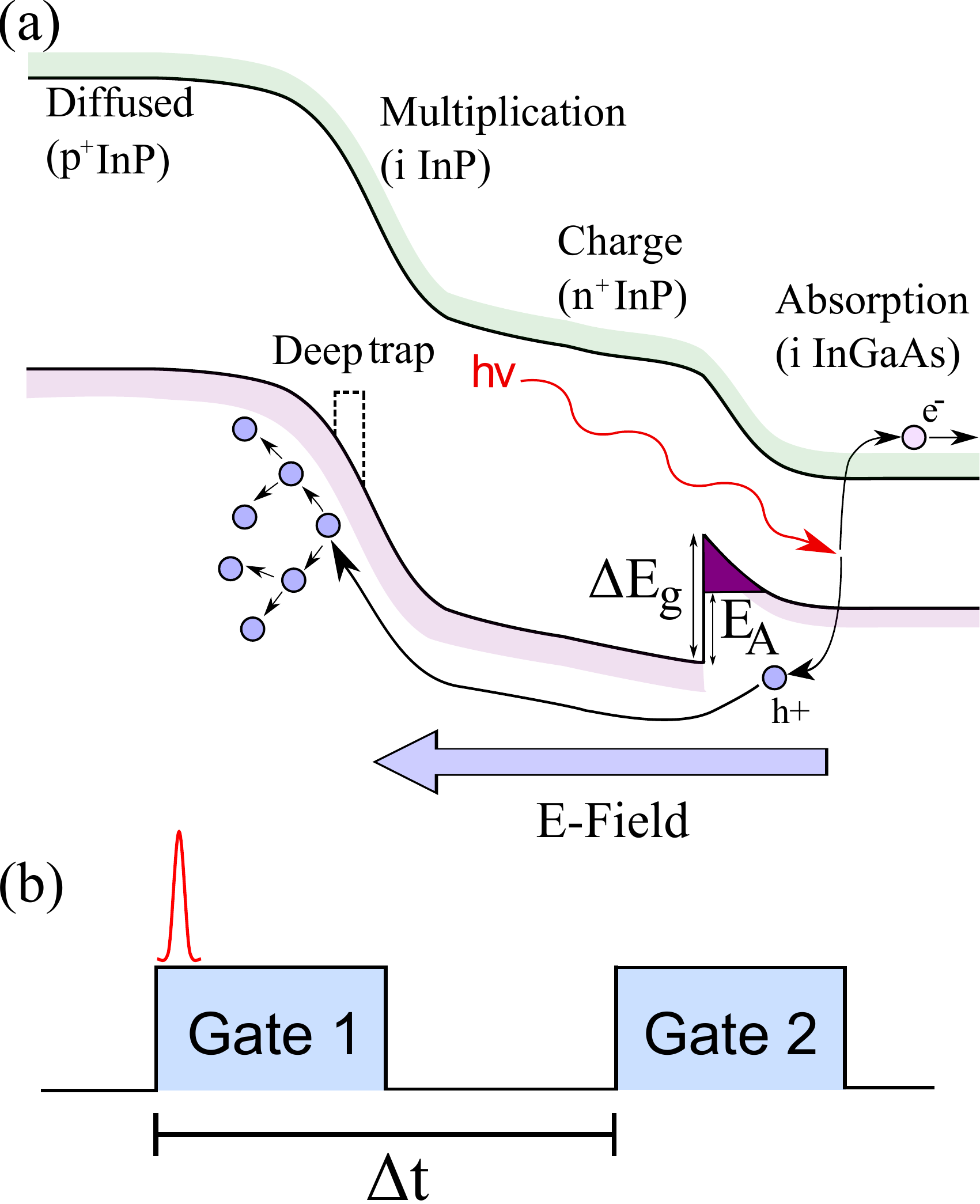}
  \caption{(a) Typical band diagram of separate absorption, charge and multiplication structure of an InGaAs/InP APD, where $E_{g}$ is the band gap offset and $E_{A}$ is the effective barrier height arising at the interface between the APD absorption and charge regions. (b) Illustration of gating scheme. Electron-hole pairs are generated at the start of Gate 1 when the laser is timed to arrival and experience an exponential decay between the two gates. The proportion of holes leftover at Gate 2 is related to the decay constant which is in turn related to the activation energy given by the barrier height, $E_{A}$.}
  \label{fig:1}
\end{figure}

To give some notion about the trapping mechanism, we provide a schematic of a typical InGaAs avalanche photodiode in Fig.~\ref{fig:1}(a). An incoming photon is absorbed in the intrinsic InGaAs region where an electron-hole pair is generated and subsequently separated by the electric field in this region.
The hole needs to overcome the potential barrier that arises from the valence-band mismatch \cite{Forrest82_APL} (the shaded purple area in Fig.~\ref{fig:1}(a)) to reach the InP multiplication region so as to have a finite probability of initiating a macroscopic avalanche which can be electronically registered.
During the generation of a macroscopic avalanche, some of the avalanche carriers may become trapped and can subsequently be released at a later time causing a secondary avalanche, known as an `afterpulse'.
The release time scale is on the order of several microseconds or greater  \cite{Jiang07,Liu07,Jiang08}.

We stress that the term `afterpulse' or `afterpulsing' only refers to clicks that are correlated with a previous detection event.
The notion of `delayed detection', on the other hand, is more general and it encompasses afterpulsing.
It refers to \textit{all} detection events originating from trapped carriers, even those that did not give rise to a detected avalanche in a previous gate.
%Therefore it encompasses both afterpulses as well as avalanches originating by other means.}

The ability of the hole to overcome the valence-band discontinuity, which is a potential barrier, directly affects device characteristics such as detection efficiency and timing response \cite{Zappa94}.
However, it is reasonable to conclude that the hole trap time is significantly shorter than 1 ns because sub-nanosecond gated-APDs still show detection efficiencies as high as 55\% \cite{comandar_gigahertz-gated_2015_aip}. If the decay time were longer than 1 ns, than fewer than half of the generated carriers would overcome the barrier and the detection efficiency would not be able to exceed 50\%.
We can infer from this analysis that the hole trap lifetime is at least 3 orders of magnitude shorter than that of deep traps causing afterpulses and %. To study the hole trap time,
we specially devised an experiment to study it, which is schematically illustrated in Fig.~\ref{fig:1}(b).
Here, we optically excite an APD at the start of a gate. When a hole fails to overcome the potential barrier within Gate 1, it will have a finite probability to overcome the barrier and initiate a macroscopic avalanche in subsequent gates within several nanoseconds.

For this study we operate the InGaAs APDs in gated Geiger mode at a clock frequency of 1 GHz. The avalanche signals are discriminated with help of self-differencing circuits that remove the capacitive response to the applied gate \cite{Yuan07}. A telecom C-band passively mode-locked laser synchronized to the APD gating frequency and with repetition frequency of 20 MHz and pulse width of 3 ps is used to illuminate the APD via its single mode fibre pigtail. We follow the best practice criteria \cite{Koehler18} to set the discrimination level of the self-differencing APD. Time-tagging electronics with a dead time of 50 ns are used to record the photon detection histogram \cite{comandar_gigahertz-gated_2015_aip}. We have measured several InGaAs APDs with different active diameters: $50 \mu m$ and $16 \mu m$. In this paper we present results from  two $16 \mu m$ devices, namely APD1 and APD2. The $50 \mu m$ devices showed similar behavior. Unless otherwise stated, the data presented is from APD~1.

\begin{figure}
  \centering
  \includegraphics[width=0.45\textwidth]{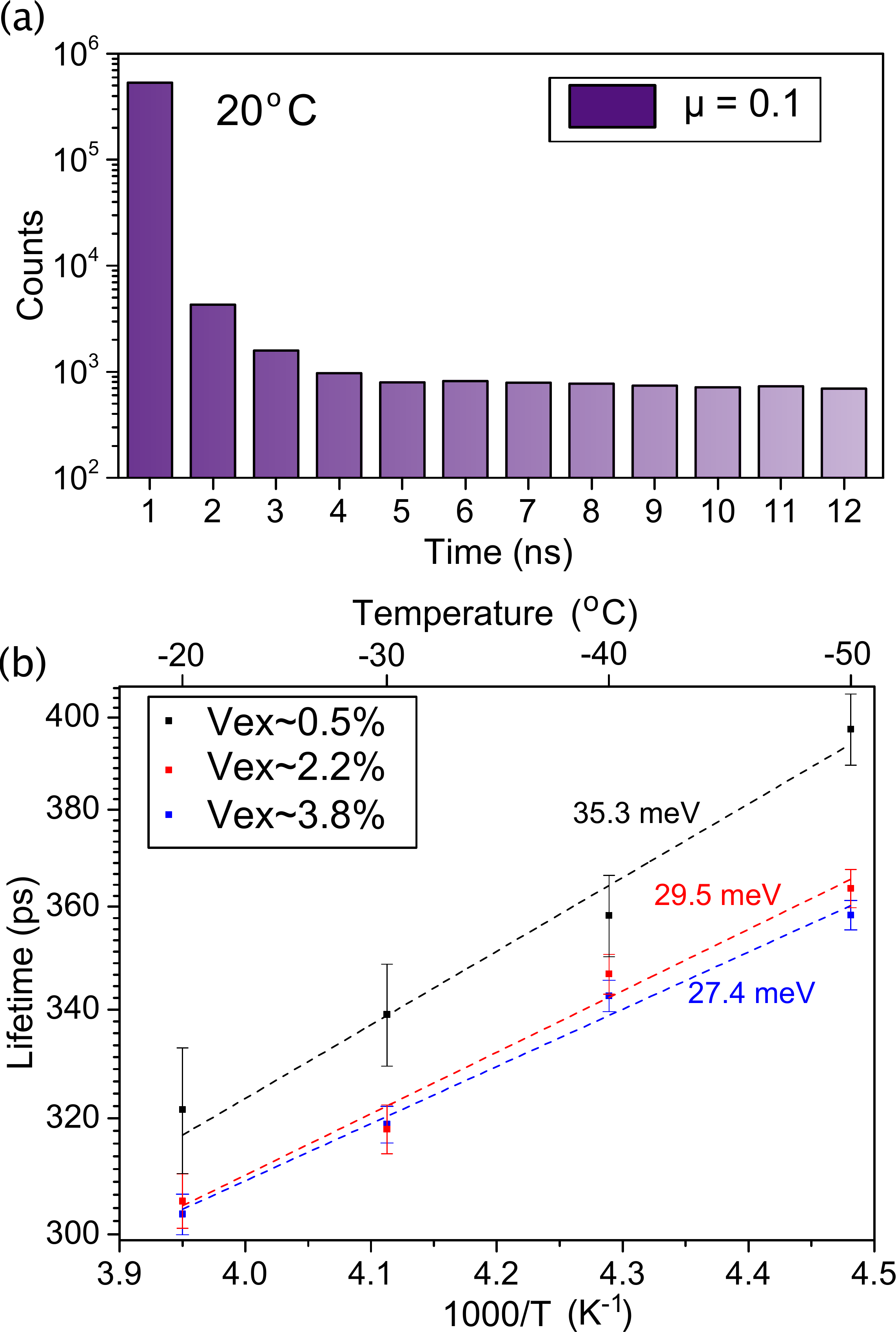}
  \caption{(a) Time-resolved histogram of detected counts of the APD under illumination of a pulsed laser with flux $\mu=0.1$, clearly demonstrating an exponential decay in counts after the initial illuminated gate; (b) An Arrhenius plot showing the single photon detection efficiency as a function of the inverse of the temperature, whereby the respective gradients allow for the extraction of the hole activation energy.}
  \label{fig:arr_plot}
\end{figure}

We first examine the role of the interface on the APD. The APD is characterised as having a single photon detection efficiency of 28\% and an afterpulsing probability of 4\% at room temperature.
Here, the optical flux is maintained at $\mu = 0.1$ photons/pulse and the laser delay is set to enable the photon arrival at the beginning of the illuminated gate (schematically shown in Fig.~\ref{fig:1}(b)), thus allowing an avalanche to have sufficient time to grow above the discrimination level and hence have a maximum detection efficiency.
Figure~\ref{fig:arr_plot}(a) shows a typical photon detection histogram under such illumination conditions. The illuminated gate gives a pronounced peak arising from single photon detections.
Immediately after this peak, the count rate experiences a fast decay before reaching an approximately flat background at the fifth gate.
The flat background is attributed to detector dark and afterpulsing counts.
The elevated count rates between 2 and 4 ns (Gates 2-4) cannot be attributed to detector afterpulsing because the time-tagger has a dead time of 50~ns.  Moreover, the sub-nanosecond decay time is orders of magnitude faster than typical lifetimes of deep traps that are responsible for afterpulsing.  We attribute the elevated count rates at these gates to delayed photon detection caused by hole trapping at the absorption/charge interface.

The above conclusion is supported by temperature dependent measurements.
It is possible to extract the interface trapping lifetime by comparing counts in Gates 1 and 3 in the histogram data (Gate 2 is neglected due to the possibility of cancellation from the self-differencer).
Plotting these lifetimes at different temperatures in an Arrhenius configuration, where the excess bias as a proportion of the breakdown voltage is kept constant for each temperature, allows for the extraction of the effective barrier height, $E_{A}$ at the material interface \cite{Forrest82_APL}, shown in Fig.~\ref{fig:arr_plot}(b), where the gradient is equal to $E_{A}/(k_{B}T)$. We note that the values of activation energies; tens of meV corresponding to lifetimes of several hundreds of picoseconds, and the trend of higher excess biases resulting in overall shorter lifetimes, and consequently lower activation energies, are consistent with the literature \cite{Forrest82_APL,Pellegrini06}. This implies that carriers with decays of several hundred picoseconds are dominated by trapping at the heterointerface when the APD is illuminated with fluxes of the order of single photons.

Carriers with the decays of several hundred picoseconds could be used to mitigate the faint after-gate attack. This is because Eve's attempt to mount such an attack using moderately high fluxes would result in delayed detection events that would alert the users to her presence. The sub-nanosecond separation between gates in GHz-clocked APDs is sufficiently narrow to allow delayed detection as a result of carriers with a decay on the order of several hundred picoseconds to be observed where they would be missed in slower, MHz-clocked systems \cite{Lydersen11,Qian18}. However, we find that in this regime, traps at the multiplication region become the dominant contribution to delayed detection events, which we now examine.

% this apparent imperfection can also be used to mitigate a class of attack available to Eve, known as the

In more detail, the after-gate attack is a class of faked state attack which itself is a type of intercept-and-resend attack \cite{makarov_*_faked_2005}. Eve measures the photons sent by the transmitter, Alice, with a copy of Bob's apparatus. She then sends her own pulses to Bob which are only detected if he chooses the same measurement basis as Eve, else he registers nothing. In this way, after Alice and Bob exchange basis information, Eve has a string that is perfectly correlated with that held by Alice and Bob. The aim for Eve is thus to send a pulse which at full power registers a click with detection probability of 1 and at half power (corresponding to incompatible bases), registers a click with probability 0. More generally, when the probability at full power exceeds twice that of half power in this manner, the detector behaviour is said to be `superlinear'. If Eve sends attack pulses towards the end of Bob's APD gate, she can maximise the ratio of detection probabilities of full and half power pulses such that she learns most of the key and also generates a sufficiently low QBER to go undetected. The original demonstration \cite{Lydersen11} involved sending pulses of moderately high photon flux ({\raise.17ex\hbox{$\scriptstyle\mathtt{\sim}$}}~40 photons/pulse) at the end of the APD gate.
%As shown in Fig.~\ref{fig:pdetgates}, using a fast-gated system means that the aforementioned time-decay constant is long enough to result in clicks in the following gate, thus signifying Eve's presence.

\begin{figure}[htbp!]
  \centering
  \includegraphics[width=0.48\textwidth]{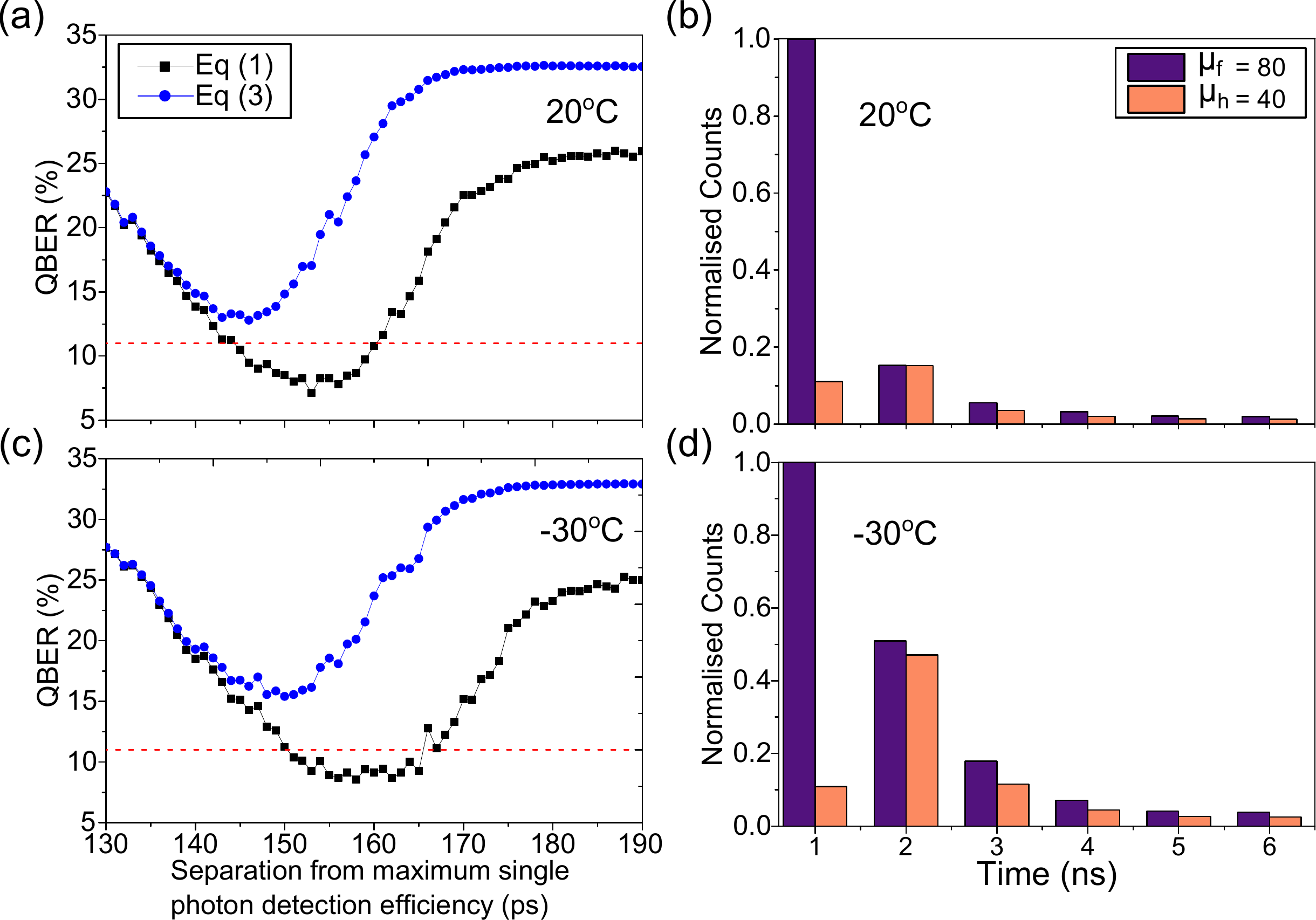}
  \caption{(a) QBER as a function of temporal separation from the maximum single photon detection efficiency delay value. The black line indicates the case where delayed detection is ignored and the QBER is calculated with equation (1) and Eve appears not to introduce a QBER greater than 11\% and thereby remains undetected. When delayed detection is taken into account, as shown in the blue line calculated with equation (2), the QBER rises above 11\% and she can be detected.; (b) Histograms taken at minimum QBER values showing detection probabilities in each Gate at 20$^{\circ}$C. Under half-power illumination of $\mu = 40$ (in orange), Gate 2 is always larger than Gate 1, which would result in a QBER value of 50\% in that Gate.; (c) as (a) but measured with the APD at --30$^{\circ}$C.; (d) as (b) but measured with the APD at --30$^{\circ}$C.}
  \label{fig:QBER}
\end{figure}

By obtaining the detection probability at full power and half power, it is possible to derive the resultant QBER using the following equation from \cite{Lydersen11}:

\begin{align}
  Q =  \frac {2p_{h} - p_{h}^2}{ 2p_{f} + 2(2p_{h} - p_{h}^2)},
  \label{eq:QBER}
\end{align}

\noindent where $p_{f}$ is the detection probability at full power and $p_{h}$ is the detection probability at half power. Note that this equation neglects any errors arising due to dark counts or afterpulsing and thus only focuses on the detection probability at the target gate. If the QBER drops below approximately 21\%, this indicates superlinearity as $p_{f} > 2p_{h}$.

We demonstrate here that GHz-gated APDs could also show superlinear behaviour when the delayed photon detections are not considered i.e. the situation when only the target gate is considered.
Here, we measured the detection probability at full (80 photons/pulse) and half (40 photons/pulse) power of an optical trigger pulse as a function of the arrival time of the laser pulse on the APD (these values were chosen due to their use in the original demonstration in \cite{Lydersen11}, however other optical powers were also investigated, the results of which are given in the Appendix).
We do this by varying the delay on the pulse generator providing the AC signal to the APD.
The result is given for two APD temperatures (20$^{\circ}$C and --30$^{\circ}$C) as the black lines in Fig. \ref{fig:QBER}(a) and (c).
At a certain temporal separation from maximum detection, the QBER drops below 11\% (illustrated as the red dotted line), reaching a minimum of approximately 7 \% at around 153 ps at room temperature, suggesting Eve could mount such an attack at this delay and remain undetected.
Either side of this trough, the QBER equals 25 \% since either $p_{f} =  p_{h} = 1$ around the centre of the gate or $p_{f} =  p_{h} \approx 0$ outside of the gate.

To probe the effect of delayed detection, we examine the histograms in the vicinity of the superlinear regime, i.e. corresponding to the conditions of an aftergate attack, as shown in Fig.~\ref{fig:QBER}(b) and (d).
We note that for the cases where Eve is using the aftergate attack, a higher proportion of clicks actually occur in the gate adjacent to the target gate (Gate 2 as opposed to Gate 1) when she chooses an incompatible basis to Bob, shown as the salmon coloured bars.
Delayed detection events would have a 50\% QBER as they are uncorrelated with Alice's qubit preparation. We note that since a higher proportion of clicks occur in the adjacent gate for incompatible bases, this would correspond to an afterpulsing probability of over 100\%, which is significantly greater than the 4\% afterpulse probability measured for the single-photon case. For compatible bases, the detection probability in Gate 2 is approximately 15\% of that in Gate 1, which is in stark contrast to the single-photon case shown in Fig.~\ref{fig:arr_plot}(a) where the Gate 2 is approximately 1\% the size of Gate 1.

We note that the degree of trapping is greater in the multi-photon case than the single-photon case for two reasons. First, more carriers are generated in the absorption region for the multi-photon case; therefore the probability of a carrier becoming ‘trapped’ at the material interface is greater. Second, as pulses are sent at the end of the gate, the electric field in the device is lower; therefore carriers that are generated and subsequently trapped in the multiplication region have a smaller probability of escaping the traps within the initial gating period and are consequently more likely to be released in the following gating period when the electric field is raised again.

This underlines the importance of incorporating delayed detection events into the calculation of the QBER. To this end, we estimate the delayed detection probabilities under full and half power pulses and add them to the detection probability without delayed detection. This leads us to the following expression for the QBER:
\begin{align}
  Q^{\prime} &=  \frac {2p_{h}^{\prime} - (p_{h}^{\prime})^2}{ 2 p_{f}^{\prime} + 2[2 p_{h}^{\prime} - (p_{h}^{\prime})^2]} ,\label{eq:QBER_dd} \\
  p_{f,h}^{\prime} &= p_{f,h} + \overline{p}_{dd} , \label{pdets_dd} \\
  \overline{p}_{dd} &= \frac{1}{4} p_{dd|f} + \frac{1}{2} p_{dd|h} \label{pdets_dd2}.
\end{align}
The quantity $Q^{\prime}$ in Eq.~\eqref{eq:QBER_dd} represents the QBER measured in the presence of the aftergate attack when delayed detection is taken into account.
This is accounted for with the term $\overline{p}_{dd}$, which represents the average probability per gate of a 1-gate-delayed detection. The factor 1/4 (1/2) in the expression is due to having a click in Bob's detectors when his basis matches (does not match) Eve's basis in the previous gate.
In Eq.~\eqref{pdets_dd2}, $p_{dd|f}$ ($p_{dd|h}$) is the probability of a delayed detection in gate $n$ when a full-power (half-power) pulse impinged on the detector at gate $n-1$, represented as a violet-coloured (salmon-coloured) bar in Fig.~\ref{fig:QBER}(b) (Fig.~\ref{fig:QBER}(d)).

Using this result, we plot the resulting QBER from Eq.~\eqref{eq:QBER_dd} with blue lines in Figs.~\ref{fig:QBER}(a) and \ref{fig:QBER}(c).
As it is apparent from the figures, the 11\% security threshold, typical of the BB84 protocol, is now overcome.
This result highlights the effectiveness of the delayed detection at mitigating the faint-after gate attack.

 By including contributions from delayed detection in Eq.~\eqref{eq:QBER_dd}, we assume Eve mounts her attack all the time.
We therefore address the case whereby Eve only attacks a fraction of gates.
In this case, the overall QBER would be smaller than the 11\% tolerance and thus Alice and Bob would not abort their key exchange.
However, Eve's information would also be smaller.
To draw a worst-case scenario, we can reason as follows \cite{lucamarini_tha_2015}. We assume for simplicity that Eve attacks "every other gate", so that she introduces errors in the odd gates and no errors in the even gates. Therefore the users can notice an odd-even pattern in the measured QBER and could draw two different key rates, one extracted from odd gates and one from even gates. The resulting key rate will be given by the sum of the two partial key rates. Due to the convexity dependence of the key rate on the QBER \cite{Gisin02,Koa06}, the resulting key rate when Eve attacks every other gate will always be larger than the one when she attacks every gate, thus confirming that it would be best for Eve to attack in every gate. This conclusion can be generalized to different attacking patterns and holds under the assumption that the users can recognize such patterns from a detailed analysis of their QBER. However we also notice that the above rationale overestimates Eve's chances to gain information because it assumes that the QBER is zero for the cases where Eve does not attack, whereas in the real case it clearly is larger than zero due to the delayed detection effect.

We also consider the case where Eve attempts to carry out a hybrid attack scheme, whereby she attempts to blind counts in Gate 2 and thus suppress any erroneous counts as a result of her after-gate attack on Gate 1.
Whilst it has been shown that blinding attacks are ineffective against appropriately operated self-differencing APDs \cite{Koehler18}, this places the onus on the user and such devices are often improperly used.
However, for Eve to blind Gate 2, due to the cancellation nature of the self-differencing circuit, she would also have to shine strong light on Gate 1, thereby negating her original attack. %Therefore such a hybrid attack would be ineffective.}

Differently to the interface trapping effect, the origin of the delay detection is predominately due to carrier trapping in the multiplication region. Consequently these delayed detection events feature longer lifetimes compared to that of interface trapping events. At 20$^{\circ}$C, the lifetimes extracted from Fig.~\ref{fig:QBER}(b) are comparable to the case shown in Fig.~\ref{fig:arr_plot}(b). However, at --30$^{\circ}$C, the lifetimes become much longer than those shown in Fig.~\ref{fig:arr_plot}(b) for the same temperature, by approximately 2-3 times. This suggests the existence of deeper traps and that these traps, rather than the material interface, are responsible for the delayed detection in the aftergate attack.  We believe these deeper traps are located in the multiplication region.

\begin{figure}
  \centering
  \includegraphics[width=0.44\textwidth]{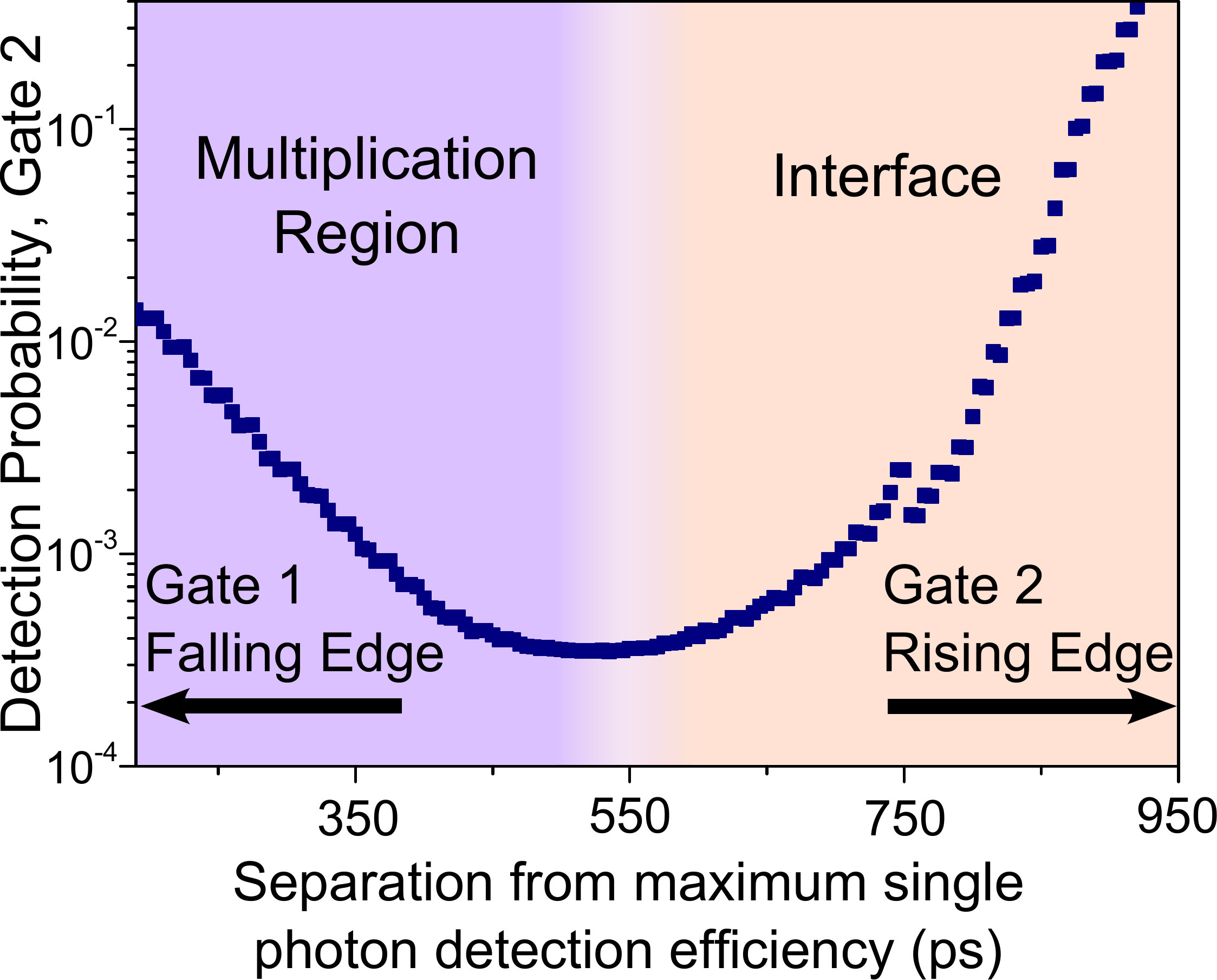}
  \caption{Detection probability in Gate 2 as a function of temporal separation from maximum single photon detection efficiency for APD~2. The increased detection probability in the left-hand side of the figure can be explained by the dominance of trapping in the multiplication region.}
  \label{fig:Pdet2}
\end{figure}

This is supported by measuring the detection probability in the adjacent gate (Gate 2 in Fig.~\ref{fig:1}(b)) as a function of separation from the maximum detection for APD~2, as shown in Fig.~\ref{fig:Pdet2}. Reading left to right, the optical photon pulse is moving away from the end of Gate 1 and approaching the start of Gate 2. The detection probability initially decreases as the laser approaches Gate 2. Here, impact ionisation is occurring and therefore carriers are multiplied and a portion of these multiplied carriers are trapped in the multiplication region, shown in purple. The high detection probability on the left-hand side roughly coincides with the QBER dip, underlining that delayed detection largely arises from trapping in the multiplication layer. If the interface was the major contributor, the detection would continue to increase the closer to Gate 2 the optical pulse is as the carriers have a progressively shorter time to decay before Gate 2 is activated. However, at a certain point the probability flattens and then begins to increase, an observation which is consistent with interface trapping, suggesting it starts to take over once carriers cease to become trapped in deep levels at the multiplication region.

Using the discovery of delayed detection allows us to define the best practice for choosing a suitable gating frequency for QKD. For this analysis at two individual temperatures, 20$^{\circ}$C and --50$^{\circ}$C, we only consider trapping at the material interface. This is the more conservative definition from a security point of view, as it requires higher gating frequencies to maintain the delayed detection required for preserving the protection against the after-gate attack. This range of gating frequencies fulfils two criteria; (i) the gating frequency is low enough to separate adjacent gates temporally such that a click in the first has a small enough probability to have a delayed detection in the second without raising the QBER above the tolerance threshold of 11\% under operation in the absence of Eve; (ii) equally, the gating frequency is high enough such that Eve would cause clicks in the gate adjacent to her target gate with a large enough probability to raise the QBER above the aforementioned threshold, which we examine for a conservative attacking flux of $\mu=20$ photons per pulse that is favourable for hiding Eve's presence (see Appendix). Our simulation result is shown in Fig.~\ref{fig:gatingf}, with the narrow white band indicating a regime where the APD is neither too `Noisy' nor `Vulnerable'. Due to the longer carrier decays at lower temperatures, we note that lower temperatures are more favourable for slower gating whereas higher temperatures are more suited to faster gating. Most significantly, gating frequencies of around 1 GHz, which are commonly used for QKD experiments (e.g. \cite{Yuan18,dixon08,Mao:18}) as well as in this study, fall in the white region, suggesting these to be optimal values for QKD.

\begin{figure}[htbp!]
  \centering
  \includegraphics[width=0.44\textwidth]{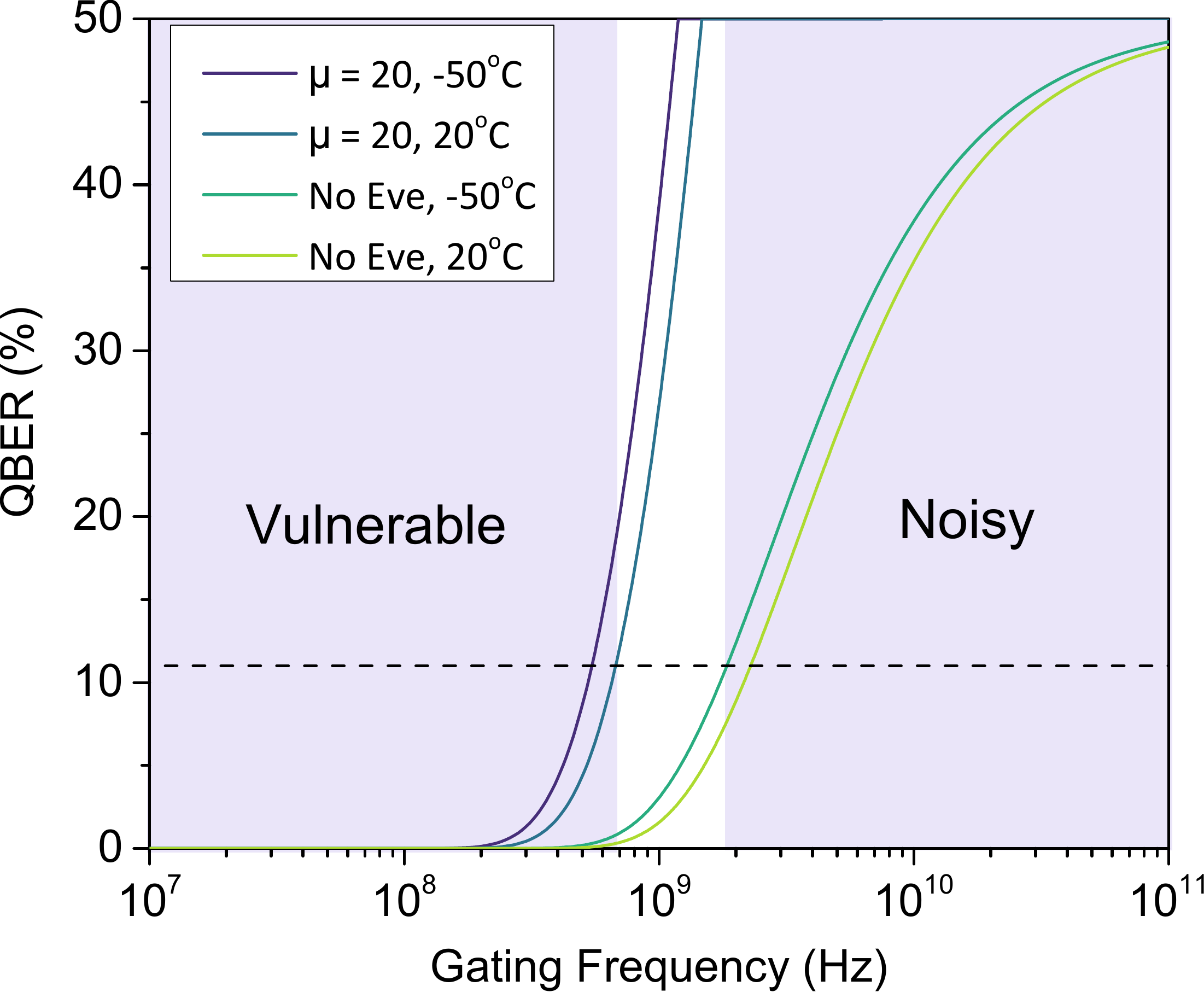}
  \caption{QBER as a function of gating frequency at 20$^{\circ}$C and --50$^{\circ}$C. The central white region indicates suitable operation, where the APD is both safe from the aftergate attack and has sufficiently low noise to make QKD possible.}
  \label{fig:gatingf}
\end{figure}

In conclusion, we have investigated two sources of trapping of carriers in InGaAs APDs: at the valence-band mismatch arising at the interface between the APD absorption and charge regions, and at deep traps in the multiplication region. In characterising the carrier lifetime at the heterojunction, we have provided the first explanation for short decays observed in fast-gated APDs. We have determined that in the after-gate regime, however, the major contribution to delayed detection events that can provide enhanced security arise from traps in the multiplication region. We have provided the first evidence that fast-gated APDs can be used to mitigate the after-gate attack due to the additional contribution to the QBER that arises from delayed detection events. By exploiting the intrinsic imperfection of the material interface, we were able to bound the appropriate APD gating frequency suitable for use in QKD.

\begin{figure}[!b]
  \centering
  \includegraphics[width=0.4\textwidth]{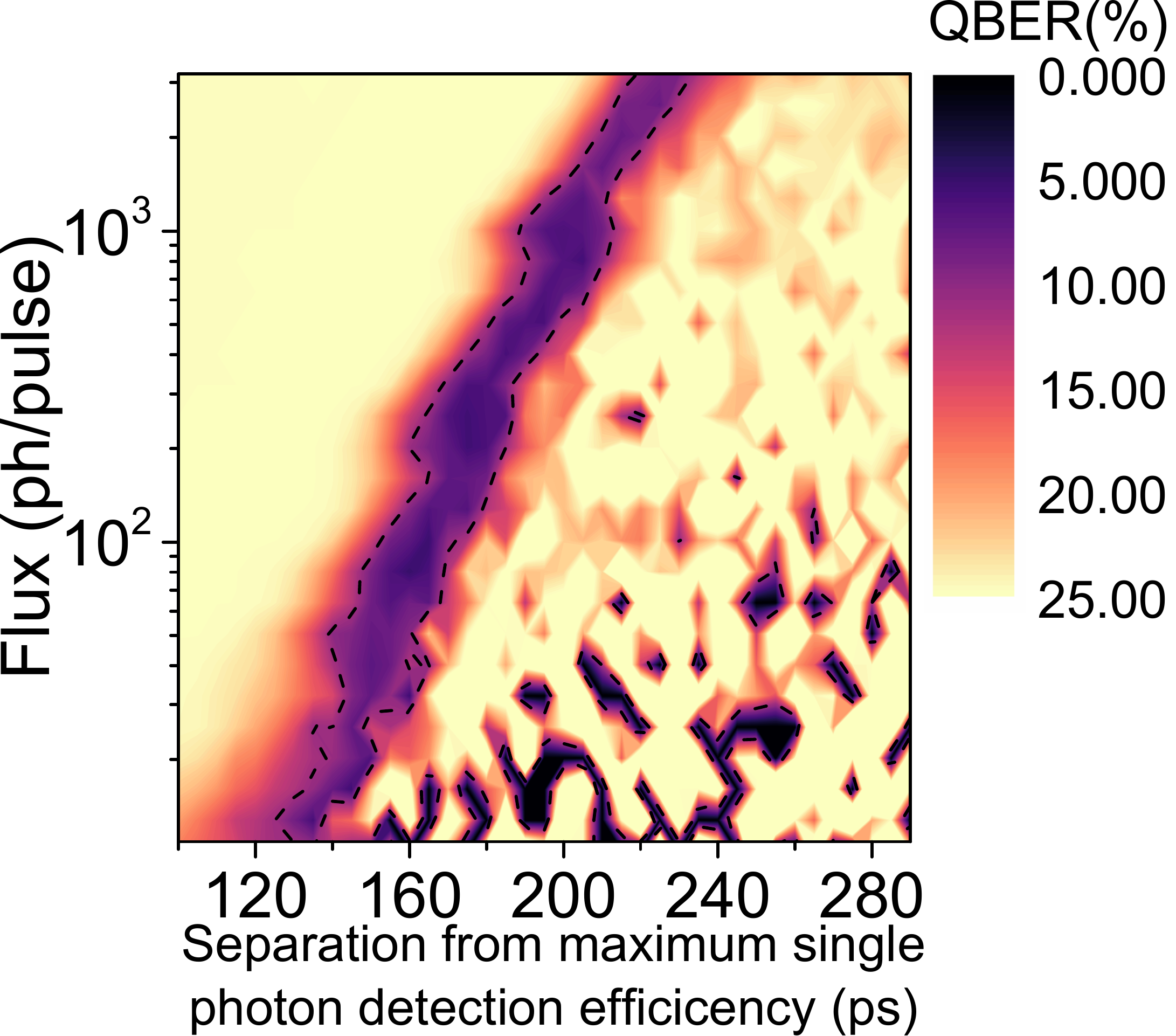}
  \caption{A contour plot of the QBER calculated using Eq.~\eqref{eq:QBER} as a function of the flux of the trigger pulse and APD gate delay with respect to the laser. The region inside the dotted line indicates where the QBER is lower than 11 \% and thus Eve can mount a successful attack in this parameter space if delayed detections are neglected.}
  \label{fig:contour}
\end{figure}

\section*{Acknowledgements}
A. K.-S. gratefully acknowledges financial support from Toshiba Research Europe Ltd and the Engineering and Physical Sciences Research Council through an Industrial CASE studentship.

\section*{Appendix}

\noindent For the demonstration of the attack presented in the paper, we chose $\mu_{f} = 80$ and $\mu_{h} = 40$ as the full and half power fluxes respectively as these were values used in the original proposal in \cite{Lydersen11}.
By expanding our measurement to examine a range of fluxes at room temperature, we were able to obtain a more general picture of the parameters that Eve could use, as shown in the measurement performed using a fast oscilloscope in Fig.~\ref{fig:contour}.

The dark purple regions within the dotted line indicate a flux and delay combination which produces a QBER that is lower than 11 \% when calculated using equation~\eqref{eq:QBER}, within which Eve will choose to operate. The pale yellow parts in the top left of the figure indicate a QBER of 25 \% which occurs when $p_{f} = p_{h} = 1$. This overall trend of this figure implies that the closer to the centre of the gate Eve moves, the smaller the flux she should use to mount her attack. This suggests that this is an extension of the original proposed after-gate attack \cite{aftergate_2011}, whereby the APD is operating in linear mode and strong pulses of power $P_{th}$ overcome the discrimination level and causes the detector to click, whereas pulses of power $P_{th}/2$ often do not and therefore rarely cause a click. By focusing on the edge of the gate, a smaller flux is required to generate the same effect which is the most favourable case for Eve. The smallest attacking flux of $\mu = 20$ was therefore used in determining appropriate gating frequencies in Fig.~\ref{fig:gatingf}.

\bibliography{aipsamp_af_changes}% Produces the bibliography via BibTeX.

\providecommand{\noopsort}[1]{}\providecommand{\singleletter}[1]{#1}%
\begin{thebibliography}{10}
\providecommand{\url}[1]{\texttt{#1}}
\providecommand{\urlprefix}{URL }
\providecommand{\eprint}[2][]{\url{#2}}

\bibitem{BB14}
C.~H. Bennett and G.~Brassard, Quantum cryptography: Public key distribution
  and coin tossing, Theor. Comput. Sci. \textbf{560}, 7 (2014).

\bibitem{peev_secoqc_2009}
M.~Peev, C.~Pacher, R.~Alléaume, C.~Barreiro, J.~Bouda, W.~Boxleitner,
  T.~Debuisschert, E.~Diamanti, M.~Dianati, J.~F. Dynes, S.~Fasel, S.~Fossier,
  M.~F\"{u}rst, J.-D. Gautier, O.~Gay, N.~Gisin, P.~Grangier, A.~Happe,
  Y.~Hasani, M.~Hentschel, H.~H\"{u}bel, G.~Humer, T.~L\"{a}nger, M.~Legr{\'e},
  R.~Lieger, J.~Lodewyck, T.~Lor\"{u}nser, N.~L\"{u}tkenhaus, A.~Marhold,
  T.~Matyus, O.~Maurhart, L.~Monat, S.~Nauerth, J.-B. Page, A.~Popper,
  E.~Querasser, G.~Ribordy, S.~Robyr, L.~Salvail, A.~W. Sharpe, A.~J. Shields,
  D.~Stucki, M.~Suda, C.~Tamas, T.~Themel, R.~T. Thew, Y.~Thoma, A.~Treiber,
  P.~Trinkler, R.~Tualle-Brouri, F.~Vannel, N.~Walenta, H.~Weier,
  H.~Weinfurter, I.~Wimberger, Z.~L. Yuan, H.~Zbinden, and A.~Zeilinger, The
  {SECOQC} quantum key distribution network in {Vienna}, New Journal of Physics
  \textbf{11}, 7, 075001 (2009).

\bibitem{sasaki11}
M.~Sasaki, M.~Fujiwara, H.~Ishizuka, W.~Klaus, K.~Wakui, M.~Takeoka, S.~Miki,
  T.~Yamashita, Z.~Wang, A.~Tanaka, K.~Yoshino, Y.~Nambu, S.~Takahashi,
  A.~Tajima, A.~Tomita, T.~Domeki, T.~Hasegawa, Y.~Sakai, H.~Kobayashi,
  T.~Asai, K.~Shimizu, T.~Tokura, T.~Tsurumaru, M.~Matsui, T.~Honjo, K.~Tamaki,
  H.~Takesue, Y.~Tokura, J.~F. Dynes, A.~R. Dixon, A.~W. Sharpe, Z.~L. Yuan,
  A.~J. Shields, S.~Uchikoga, M.~Legr\'{e}, S.~Robyr, P.~Trinkler, L.~Monat,
  J.-B. Page, G.~Ribordy, A.~Poppe, A.~Allacher, O.~Maurhart, T.~L\"{a}nger,
  M.~Peev, and A.~Zeilinger, Field test of quantum key distribution in the
  {Tokyo} {QKD} {Network}, Opt. Express \textbf{19}, 11, 10387 (2011).

\bibitem{Dynes12}
J.~F. Dynes, I.~Choi, A.~W. Sharpe, A.~R. Dixon, Z.~L. Yuan, M.~Fujiwara,
  M.~Sasaki, and A.~J. Shields, Stability of high bit rate quantum key
  distribution on installed fiber, Opt. Express \textbf{20}, 15, 16339 (2012).

\bibitem{Mao:18}
Y.~Mao, B.-X. Wang, C.~Zhao, G.~Wang, R.~Wang, H.~Wang, F.~Zhou, J.~Nie,
  Q.~Chen, Y.~Zhao, Q.~Zhang, J.~Zhang, T.-Y. Chen, and J.-W. Pan, Integrating
  quantum key distribution with classical communications in backbone fiber
  network, Opt. Express \textbf{26}, 5, 6010 (2018).

\bibitem{dixon_2017}
A.~R. Dixon, J.~F. Dynes, M.~Lucamarini, B.~Fr\"{o}hlich, A.~W. Sharpe,
  A.~Plews, W.~Tam, Z.~L. Yuan, Y.~Tanizawa, H.~Sato, S.~Kawamura, M.~Fujiwara,
  M.~Sasaki, and A.~J. Shields, Quantum key distribution with hacking
  countermeasures and long term field trial, Sci. Rep. \textbf{7}, 1978 (2017).

\bibitem{Sun18}
W.~Sun, L.-J. Wang, X.-X. Sun, Y.~Mao, H.-L. Yin, B.-X. Wang, T.-Y. Chen, and
  J.-W. Pan, Experimental integration of quantum key distribution and
  gigabit-capable passive optical network, Journal of Applied Physics
  \textbf{123}, 4, 043105 (2018).

\bibitem{Bunandar18}
D.~Bunandar, A.~Lentine, C.~Lee, H.~Cai, C.~M. Long, N.~Boynton, N.~Martinez,
  C.~DeRose, C.~Chen, M.~Grein, D.~Trotter, A.~Starbuck, A.~Pomerene,
  S.~Hamilton, F.~N.~C. Wong, R.~Camacho, P.~Davids, J.~Urayama, and
  D.~Englund, Metropolitan quantum key distribution with silicon photonics,
  Phys. Rev. X \textbf{8}, 021009 (2018).

\bibitem{makarov_*_faked_2005}
V.~Makarov and D.~R. Hjelme, Faked states attack on quantum cryptosystems, J.
  Mod. Opt. \textbf{52}, 5, 691 (2005).

\bibitem{sauge_controlling_2011}
S.~Sauge, L.~Lydersen, A.~Anisimov, J.~Skaar, and V.~Makarov, Controlling an
  actively-quenched single photon detector with bright light, Opt. Express
  \textbf{19}, 23, 23590 (2011).

\bibitem{vakhitov_tha_2011}
A.~Vakhitov, V.~Makarov, and D.~R. Hjelme, Large pulse attack as a method of
  conventional optical eavesdropping in quantum cryptography, J. Mod. Opt.
  \textbf{48}, 13, 2023 (2001).

\bibitem{gisin_tha_2006}
N.~Gisin, S.~Fasel, B.~Kraus, H.~Zbinden, and G.~Ribordy, Trojan-horse attacks
  on quantum-key-distribution systems, Phys. Rev. A \textbf{73}, 022320 (2006).

\bibitem{jiang_intrinsic_2013}
M.-S. Jiang, S.-H. Sun, G.-Z. Tang, X.-C. Ma, C.-Y. Li, and L.-M. Liang,
  Intrinsic imperfection of self-differencing single-photon detectors harms the
  security of high-speed quantum cryptography systems, Phys. Rev. A
  \textbf{88}, 6, 062335 (2013).

\bibitem{yuan2010avoiding}
Z.~Yuan, J.~Dynes, and A.~Shields, Avoiding the blinding attack in {QKD}, Nat.
  Photon. \textbf{4}, 12, 800 (2010).

\bibitem{yuan_resilience_2011}
Z.~L. Yuan, J.~F. Dynes, and A.~J. Shields, Resilience of gated avalanche
  photodiodes against bright illumination attacks in quantum cryptography,
  Appl. Phys. Lett. \textbf{98}, 23, 231104 (2011).

\bibitem{Lydersen_bitmap_11}
L.~Lydersen, V.~Makarov, and J.~Skaar, Secure gated detection scheme for
  quantum cryptography, Phys. Rev. A \textbf{83}, 032306 (2011).

\bibitem{lucamarini_tha_2015}
M.~Lucamarini, I.~Choi, M.~B. Ward, J.~F. Dynes, Z.~L. Yuan, and A.~J. Shields,
  {Practical Security Bounds Against the Trojan-Horse Attack in Quantum Key
  Distribution}, Phys. Rev. X \textbf{5}, 031030 (2015).

\bibitem{Lim15}
C.~C.~W. {Lim}, N.~{Walenta}, M.~{Legré}, N.~{Gisin}, and H.~{Zbinden}, Random
  variation of detector efficiency: A countermeasure against detector blinding
  attacks for quantum key distribution, IEEE Journal of Selected Topics in
  Quantum Electronics \textbf{21}, 3, 192 (2015).

\bibitem{daSilva15}
T.~{Ferreira da Silva}, G.~C. {do Amaral}, G.~B. {Xavier}, G.~P. {Temporão},
  and J.~P. {von der Weid}, Safeguarding quantum key distribution through
  detection randomization, IEEE Journal of Selected Topics in Quantum
  Electronics  (2015).

\bibitem{Koehler18}
A.~Koehler-Sidki, J.~F. Dynes, M.~Lucamarini, G.~L. Roberts, A.~W. Sharpe,
  Z.~L. Yuan, and A.~J. Shields, {Best-Practice Criteria for Practical Security
  of Self-Differencing Avalanche Photodiode Detectors in Quantum Key
  Distribution}, Physical Review Applied \textbf{9}, 044027 (2018).

\bibitem{KoehlerSidkiIM18}
A.~Koehler-Sidki, M.~Lucamarini, J.~F. Dynes, G.~L. Roberts, A.~W. Sharpe,
  Z.~Yuan, and A.~J. Shields, Intensity modulation as a preemptive measure
  against blinding of single-photon detectors based on self-differencing
  cancellation, Phys. Rev. A \textbf{98}, 022327 (2018).

\bibitem{lydersen_hacking_2010}
L.~Lydersen, C.~Wiechers, C.~Wittmann, D.~Elser, J.~Skaar, and V.~Makarov,
  Hacking commercial quantum cryptography systems by tailored bright
  illumination, Nat. Photon. \textbf{4}, 10, 686 (2010).

\bibitem{Makarov16}
V.~Makarov, J.-P. Bourgoin, P.~Chaiwongkhot, M.~Gagn\'e, T.~Jennewein,
  S.~Kaiser, R.~Kashyap, M.~Legr\'e, C.~Minshull, and S.~Sajeed, Creation of
  backdoors in quantum communications via laser damage, Phys. Rev. A
  \textbf{94}, 030302(R) (2016).

\bibitem{Pinheiro18}
P.~V.~P. Pinheiro, P.~Chaiwongkhot, S.~Sajeed, R.~T. Horn, J.-P. Bourgoin,
  T.~Jennewein, N.~L\"{u}tkenhaus, and V.~Makarov, Eavesdropping and
  countermeasures for backflash side channel in quantum
  cryptography\color{black}, Opt. Express \textbf{26}, 16, 21020 (2018).

\bibitem{gerhardt_full-field_2011}
I.~Gerhardt, Q.~Liu, A.~Lamas-Linares, J.~Skaar, C.~Kurtsiefer, and V.~Makarov,
  Full-field implementation of a perfect eavesdropper on a quantum cryptography
  system, Nature Communications \textbf{2}, 349 (2011).

\bibitem{lo12}
H.-K. Lo, M.~Curty, and B.~Qi, Measurement-device-independent quantum key
  distribution, Phys. Rev. Lett. \textbf{108}, 13, 130503 (2012).

\bibitem{Braunstein12}
S.~L. Braunstein and S.~Pirandola, Side-channel-free quantum key distribution,
  Phys. Rev. Lett. \textbf{108}, 130502 (2012).

\bibitem{LYDS18}
M.~Lucamarini, Z.~L. Yuan, J.~F. Dynes, and A.~J. Shields, Overcoming the
  {rate–-distance} limit of quantum key distribution without quantum
  repeaters, Nature \textbf{557}, 400 (2018).

\bibitem{comandar_gigahertz-gated_2015_aip}
L.~C. Comandar, B.~Fr\"{o}hlich, J.~F. Dynes, A.~W. Sharpe, M.~Lucamarini,
  Z.~L. Yuan, R.~V. Penty, and A.~J. Shields, Gigahertz-gated {InGaAs}/{InP}
  single-photon detector with detection efficiency exceeding 55\% at 1550 nm,
  J. Appl. Phys. \textbf{117}, 8, 083109 (2015).

\bibitem{comandar_rmtemp_2014}
L.~C. Comandar, B.~Fr\"{o}hlich, M.~Lucamarini, K.~A. Patel, A.~W. Sharpe,
  J.~F. Dynes, Z.~L. Yuan, R.~V. Penty, and A.~J. Shields, Room temperature
  single-photon detectors for high bit rate quantum key distribution, Appl.
  Phys. Lett. \textbf{104}, 2, 021101 (2014).

\bibitem{Yuan18}
Z.~Yuan, A.~Plews, R.~Takahashi, K.~Doi, W.~Tam, A.~Sharpe, A.~Dixon,
  E.~Lavelle, J.~Dynes, A.~Murakami, M.~Kujiraoka, M.~Lucamarini, Y.~Tanizawa,
  H.~Sato, and A.~J. Shields, {10-Mb/s Quantum Key Distribution}, Journal of
  Lightwave Technology \textbf{36}, 16, 3427 (2018).

\bibitem{Boaron18}
A.~Boaron, B.~Korzh, R.~Houlmann, G.~Boso, D.~Rusca, S.~Gray, M.-J. Li,
  D.~Nolan, A.~Martin, and H.~Zbinden, Simple {2.5~GHz} {time--bin} quantum key
  distribution, Applied Physics Letters \textbf{112}, 17, 171108 (2018).

\bibitem{Lydersen11}
L.~Lydersen, N.~Jain, C.~Wittmann, O.~Mar\o{}y, J.~Skaar, C.~Marquardt,
  V.~Makarov, and G.~Leuchs, Superlinear threshold detectors in quantum
  cryptography, Phys. Rev. A \textbf{84}, 032320 (2011).

\bibitem{Qian18}
Y.-J. Qian, D.-Y. He, S.~Wang, W.~Chen, Z.-Q. Yin, G.-C. Guo, and Z.-F. Han,
  Hacking the quantum key distribution system by exploiting the
  avalanche-transition region of single-photon detectors, Phys. Rev. Applied
  \textbf{10}, 064062 (2018).

\bibitem{Forrest82_APL}
S.~R. Forrest, O.~K. Kim, and R.~G. Smith, Optical response time of
  {In0.53Ga0.47As/InP} avalanche photodiodes, Applied Physics Letters
  \textbf{41}, 1, 95 (1982).

\bibitem{Jiang07}
X.~Jiang, M.~A. Itzler, R.~Ben-Michael, and K.~Slomkowski, {InGaAsP–InP
  Avalanche Photodiodes for Single Photon Detection}, IEEE Journal of Selected
  Topics in Quantum Electronics \textbf{13}, 4, 895 (2007).

\bibitem{Liu07}
M.~Liu, C.~Hu, X.~Bai, X.~Guo, J.~C. Campbell, Z.~Pan, and M.~M. Tashima,
  {High-Performance} {InGaAs/InP} {Single-Photon} {Avalanche} {Photodiode},
  IEEE Journal of Selected Topics in Quantum Electronics \textbf{13}, 4, 887
  (2007).

\bibitem{Jiang08}
X.~Jiang, M.~A. Itzler, R.~Ben-Michael, K.~Slomkowski, M.~A. Krainak, S.~Wu,
  and X.~Sun, {Afterpulsing Effects in Free-Running InGaAsP Single-Photon
  Avalanche Diodes}, IEEE Journal of Quantum Electronics \textbf{44}, 1, 3
  (2008).

\bibitem{Zappa94}
F.~Zappa, A.~Lacaita, S.~Cova, and P.~Webb, {Nanosecond single-photon timing
  with InGaAs/InP photodiodes}, Opt. Lett. \textbf{19}, 11, 846 (1994).

\bibitem{Yuan07}
Z.~L. Yuan, B.~E. Kardynal, A.~W. Sharpe, and A.~J. Shields, High speed single
  photon detection in the near infrared, Appl. Phys. Lett. \textbf{91}, 4,
  041114 (2007).

\bibitem{Pellegrini06}
S.~Pellegrini, R.~E. Warburton, L.~J.~J. Tan, J.~S. Ng, A.~B. Krysa, K.~Groom,
  J.~P.~R. David, S.~Cova, M.~J. Robertson, and G.~S. Buller, {Design and
  performance of an InGaAs-InP single-photon avalanche diode detector}, IEEE
  Journal of Quantum Electronics \textbf{42}, 4, 397 (2006).

\bibitem{Gisin02}
N.~Gisin, G.~Ribordy, W.~Tittel, and H.~Zbinden, Quantum cryptography, Rev.
  Mod. Phys. \textbf{74}, 145 (2002).

\bibitem{Koa06}
M.~Koashi, Efficient quantum key distribution with practical sources and
  detectors (2006), \eprint{arXiv:quant-ph/0609180}.

\bibitem{dixon08}
A.~R. Dixon, Z.~L. Yuan, J.~F. Dynes, A.~W. Sharpe, and A.~J. Shields,
  Gigahertz decoy quantum key distribution with 1 \uppercase{M}bit/s secure key
  rate, Opt. Express \textbf{16}, 23, 18790 (2008).

\bibitem{aftergate_2011}
C.~Wiechers, L.~Lydersen, C.~Wittmann, D.~Elser, J.~Skaar, C.~Marquardt,
  V.~Makarov, and G.~Leuchs, After-gate attack on a quantum cryptosystem, New
  J. Phys. \textbf{13}, 1, 013043 (2011).

\end{thebibliography}
\bibliographystyle{mybst}

\end{document}